\documentclass[prb,twocolumn,showpacs,amsmath,amssymb]{revtex4}

\usepackage{graphicx}% Include figure files
\usepackage{dcolumn}% Align table columns on decimal point
\usepackage{bm}% bold math

\begin{document}

%\preprint{APS/123-QED}

\title{
  Polar behavior of double perovskite (Bi,Pb)ZnNbO$_6$ and
  (Bi,Sr)ZnNbO$_6$: Density functional calculations
}

\author{Shigeyuki Takagi$^{1,2}$}
\author{Alaska Subedi$^{1,2}$}
\author{David J. Singh$^{1}$}
\author{Valentino R. Cooper$^{1}$}
%\email{takagism@ornl.gov}
\affiliation{
  $^1$Materials Science and Technology Division,
Oak Ridge National Laboratory,
  Oak Ridge, TN 37831-6114 \\
  $^2$Department of Physics, University of Tennessee, Knoxville,
  TN 37996
}

\date{\today}

\begin{abstract}
  The polar behavior of double perovskite (Bi,Pb)ZnNbO$_6$ and
  (Bi,Sr)ZnNbO$_6$ was investigated using first principles
  density functional
calculations within the local density approximation.
These materials have both $A$-site size disorder and
ions with sterochemical activity.
  We found a strong ferroelectric distortion in both materials.
  The polarization of the Pb material is $>$80 $\mu$C/cm$^2$ along the
  rhombohedral direction and the Sr
based material is only slightly inferior.
The high polarization results mainly
from a large off-centering of Bi and the large Born effective charge
of Nb, which yields a large contribution polarization although the Nb 
off-centers by a smaller amount.
 Neither of these materials favors a
  tetragonal state, and therefore solid solutions
 with PbTiO$_3$ may show morphotoropic
  phase boundaries.
\end{abstract}


\maketitle

\section{INTRODUCTION}

Perovskite ferroelectrics and relaxor ferroelectrics
are widely used in applications including electromechanical
transducers, actuators, passive electronic components,
memory devices and microwave technology.
There is strong interest in obtaining improvements in materials performance,
such as high polarization, high Curie temperature,
improved electromechanical coupling and lower loss.
One approach is through the exploration of new materials.
Besides this technological motivation, these materials have
been the focus of considerable fundamental interest,
particularly in unraveling the complex interplay between crystal
chemistry and lattice distortions and in understanding the range of polar
behaviors that can be produced by strain and chemical substitutions.
Much has been learned over the past several years.
  
The lattice instabilities of perovskite oxides, $AB\text{O}_3$ are often
understood using the tolerance factor
$t=(r_{\text{O}}+r_A)/\sqrt{2}(r_{\text{O}}+r_B)$, where $r_{\text{O}}$,
$r_A$ and $r_B$ are the ionic radii for the O, $A$-site and $B$-site ions,
respectively.\cite{goldschmidt,shannon1976acta,zhong1995prl}
These materials are classed in two groups
according to whether the
tolerance factor is less than unity or not.
Several $t>1$ ferroelectrics are known, including the
prototypical materials, BaTiO$_3$ and KNbO$_3$.
In these materials there is a tendency towards lattice distortions
where the too-small $B$-site ion off-centers. This is strongly enhanced
by covalency involving O $2p$ states and nominally unoccupied
$B$-site $d$ states. This then leads to enhanced Born effective charges and
large longitudinal optic - transverse optic splittings.
\cite{cohen1992nature,zhong}
  In the majority of $t<1$ materials the $B\text{O}_6$
  octahedra are tilted and they are not ferroelectrics.\cite{choi2005jap}
  The exceptions are generally perovskites
with lone-pair $A$-sites such as Pb and Bi.
In these cases, the ideal cubic perovskite structure is often
unstable against octahedral tilting, but the ferroelectric instability,
which is enhanced by the stereochemical activity of the $A$-site, is stronger.
In materials with $t<1$ but no stereochemically active $A$-site ion,
there is often a substantial
ferroelectric instability of the ideal cubic
structure, but now the tilt instability is stronger, and so
ferroelectricity does not occur.
\cite{halilov2002apl,bilc2006prl,singh2006ferroelectrics}
However, according to
first principles calculations,
if the octahedra are prevented from tilting by mixing large and
small $A$-site ions they may become
ferroelectrics.\cite{singh2008prl}
There is experimental support for this. For example, ferroelectricity
in the (Ba,Ca)TiO$_3$ solid solution
is strongly enhanced compared to what would
be expected interpolating between the properties of the end-point
compounds. \cite{fu-bct}
Importantly, this leads to new compositions that have high piezoelectric
coefficients. \cite{fu-bct,liu-bct}
 
Generally, high piezoelectric coefficients and electromechanical
coupling are found near phase boundaries. In particular,
the electromechanical coupling in perovskite piezoelectrics is
associated with polarization rotation near morphotropic phase
boundaries (MPBs).
\cite{noheda,fu-cohen}
These are generally boundaries between ferroelectric states differing
in the direction of their ground state polarization.
Therefore one strategy for finding new piezoelectric compositions
is to search for solid solutions with MPBs.
A particular emphasis has been in solid solutions with PbTiO$_3$ (PT)
as one end-point, following the commonly used
Pb(Zr,Ti)O$_3$ (PZT) system and the PbMg$_{1/3}$Nb$_{2/3}$O$_3$ (PMN) - PT
and PbZn$_{1/3}$Nb$_{2/3}$O$_3$ (PZN) - PT piezocrystal systems.
\cite{shrout-pc}
PT is tetragonal with a substantial tetragonality.
There is considerable
interest in finding materials that can be alloyed with PT
and which have strong ferroelectricity with rhombohedral or related
ground states to produce an MPB perhaps with high Curie temperature,
high polarization and high tetragonality on the tetragonal side of the
MPB.

A development of potential importance is the realization that Bi based
perovskites can be very useful ferroelectrics, as exemplified by
progress in BiFeO$_3$. In particular, the small ionic radius
and stereochemical activity of Bi$^{3+}$,
leads to large displacements
in a low volume unit cell, and its high charge (3+ vs. 2+ for Pb)
and hybridization 
of Bi 6$p$ states with O 2$p$ states leads to a high Born charge.
\cite{neaton}
Furthermore,
experimental work on the BiScO$_3$ - PT solid solution
shows an MPB with enhanced Curie temperature, \cite{eitel}
and there have been recent studies of several other Bi-based perovskites
in solid solutions with PT.
\cite{stringer2005jap,suchomel2004jap,randall2004jap,cheng2003materlett,
duan2004jmaterres}.
One problem has been that while the small Bi-ion tends to enhance
tetragonality on the tetragonal side of the MPB,
this same effect may prevent the occurrence of an MPB, leading to
phase diagrams with only tetragonal phases.

Recently, an extremely large tetragonality
($c/a \sim$ 1.11 and high Curie temperature
  $T_\text{C}$ of $\sim$ 700 ${}^\circ$C at $x=0.6$
was reported for perovskite solution
  ($1-x$)Bi(Zn$_{1/2}$Ti$_{1/2}$)O$_3$-($x$)PbTiO$_3$.
\cite{suchomel2005apl}
This is reminiscent of the behavior of CdTiO$_3$ - PT alloys,
which also have highly enhanced tetragonality.
\cite{halilov-pct,suarez-sandoval,chen}
In perovskites the balance between tetragonal and rhombohedral
states is generally controlled by a balance between the energy
lowering due to tetragonal strain, which favors the tetragonal
state, and the $B$-site off-centering, which favors a rhombohedral
state \cite{ghita}
(note that with small $A$-site ions there is an additional stabilization
of the tetragonal state because
off-centerings along [001] directions are towards the most
open face of the O cage around the $A$-site;
therefore, this direction is favored for the $A$-site and this can
be important if the displacement is very large
\cite{bilc2006prl}).
The interplay of $A$-site
and $B$-site off-centering, and the importance of the $B$-site
in determining the direction of the ferroelectric polarization
suggests the exploration of alternate $B$-site ions in solid solutions with
PT to obtain an MPBs.

Here, we present a study of the polar behavior of perovskite
(Bi,Pb)(Zn,Nb)O$_6$ and (Bi,Sr)(Zn,Nb)O$_6$ using density functional
supercell calculations.
The purpose is to explore the effect of combining the mechanisms
discussed above in a perovskite that may be amenable to experimental
synthesis. Specifically, the compositions explored have different
size $A$-site cations (Pb and Bi), stereochemically active ions on the
$A$-site (particularly, Bi, though Pb may also contribute as discussed
below), and
ions with electronic structures that favor cross-gap hybridization and
therefore ferroelectricity on the $B$-site (Nb and Zn).

As mentioned, one motivation for this study is to explore
the use of different size $A$-site ions in relation to stereochemical activity
to obtain $A$-site driven ferroelectricity.
The ionic radii \cite{shannon1976acta}
of the $A$-site ions considered here are
$r_{\text{Pb}^{2+}}=1.63$ \AA, $r_{\text{Sr}^{2+}}=1.58$ \AA,
and $r_{\text{Bi}^{3+}}=1.31$ \AA.
Thus, Bi$^{3+}$ is significantly smaller than the other $A$-site ions
and Pb and Sr have approximately the same size, but unlike Sr, Pb
has stereochemical activity that can favor ferroelectricity, as in PbTiO$_3$.

Related to this there is an experimental report of the synthesis and
some physical properties of perovskite
(Bi$_{0.5}$Sr$_{0.5}$)(Mg$_{0.5}$Nb$_{0.5}$)O$_3$,
(Bi$_{0.5}$Ba$_{0.5}$)(Mg$_{0.5}$Nb$_{0.5}$)O$_3$,
(Bi$_{0.5}$Sr$_{0.5}$)(Zn$_{0.5}$Nb$_{0.5}$)O$_3$,
(Bi$_{0.5}$Ba$_{0.5}$)(Zn$_{0.5}$Nb$_{0.5}$)O$_3$, and the
solid solution of (Bi$_{0.5}$Ba$_{0.5}$)(Zn$_{0.5}$Nb$_{0.5}$)O$_3$
with PT and
(Bi$_{0.5}$Sr$_{0.5}$)(Mg$_{0.5}$Nb$_{0.5}$)O$_3$ with SrTiO$_3$.
\cite{bogatko,kosyachenko}
The compound (Bi$_{0.5}$Sr$_{0.5}$)(Mg$_{0.5}$Nb$_{0.5}$)O$_3$
was reported to be ferroelectric based on the observation of
a hysteresis loop below 103 K.
All four compounds were reported to show phase transitions
above room temperature and relatively high dielectric constants.
They also show
substantial microwave loss, which may indicate ferroelectricity.

\section{Approach}

The perovskite alloys were studied using supercell calculations
within density functional theory. The main results are for
compositions BiPbNbZnO$_6$ and BiSrNbZnO$_6$.
The large charge difference between Zn$^{2+}$ and Nb$^{5+}$
may be expected to lead to a strong ordering tendency on the
$B$-site. This is the case in the relaxor systems
PbZn$_{1/3}$Nb$_{2/3}$O$_3$ (PZN) and PbMg$_{1/3}$Nb$_{2/3}$O$_3$ (PMN),
where a double perovskite like structure occurs even though
the Nb:Zn stoichiometry is not 1:1
(specifically, in PZN and PMN one double perovskite $B$-site sublattice is Nb 
and the other is 1/3 Nb and 2/3 Zn or Mg, which in turn may further order).
\cite{krause,davies-pmn,prosandeev}
In the present case the stoichiometry is 1:1, which should further
stabilize the ordering.
Therefore we assume an ordering of the $B$-site lattice into a double
perovskite structure.
In contrast, there is no a priori reason to assume that the $A$-site lattice
consisting of Pb and Bi is chemically ordered.

BiPbZnNbO$_6$ (BPZN) and BiSrZnNbO$_6$ (BSZN) have
ferroelectrically active ions on both the $A$ and $B$ sites.
On the $B$-site both Zn$^{2+}$ and Nb$^{5+}$ favor ferroelectricity,
as is evident from the stronger relaxor ferroelectricity of PZN relative
to PMN and the ferroelectricity of KNbO$_3$.
Likewise Bi$^{3+}$ and Pb$^{2+}$ are stereochemically active, and favor
ferroelectricity on the perovskite $A$-site.
However, both BPZN and BSZN
are expected to be strongly
$A$-site driven materials both based on ionic radius ($t < 1$, and
$A$-site size disorder) and
lone pair physics, \cite{ghita}
and as such a competition between octahedral tilts and polar off-centering
may be expected similar to the Pb(Zr,Ti)O$_3$ system. \cite{fornari-tilt}
Therefore it is important that the supercells are constructed in a way that
this competition can be included. The minimum size for this is 2x2x2,
or 40 atoms, since such a cell has an even number of units along
the [001], [011], and [111] directions and therefore can accommodate
the various Glazer tilt patterns. \cite{glazer}

An additional constraint comes from the fact that with multiple
$A$-site ions it is possible to choose cells that have polar spacegroups
just because of the cation ordering and not because of lattice instability.
This would not represent ferroelectricity, but would rather simply be an
artifact of the selected order.
Therefore even though Bi and Pb are not expected to chemically order
in the solid solutions, we consider them in a highly ordered state
within our supercells. Here we select a rock-salt ordering of Pb and Bi
on the $A$-site. The cation ordering with this choice has symmetry
$F\bar{4}3m$, which is non-polar. Thus any polarization or off-centering
is a consequence of lattice instability and not the choice of cation
ordering, and furthermore, one may expect that a more disordered alloy
would have a better frustration of the tilts and a stronger ferroelectric
tendency. \cite{singh2008prl}

The density functional (DFT) calculations were done using two methods.
We used the general potential linearized augmented planewave (LAPW) method
\cite{singh-book}
for the structure relaxations. This is a full potential all electron
method. We did the calculations using the local density approximation
at a scalar relativistic level.
Well converged basis sets were employed with LAPW sphere radii of
2.3 Bohr for Pb, Sr and Bi, 1.90 Bohr for Zn and Nb and 1.55 Bohr for O.
We included local orbitals to relax linearization and to accurately treat
semicore states. \cite{singh-lo}
The polarization and Born effective
charges were calculated with an ultrasoft pseudopotential method
as implemented in the quantum espresso package, \cite{qe}
using the relaxed structures from the LAPW calculations.
The results were converged with respect to the Brillouin zone sampling,
which was tested.
The polarization was obtained using the Berry's phase method.

\section{RESULTS}

\begin{figure}
\includegraphics[width=\columnwidth,angle=0]{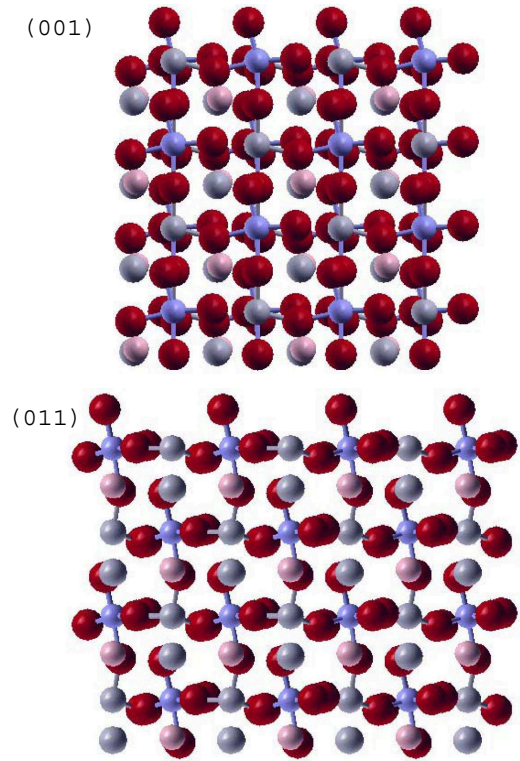}
\caption{(color online) Structure of the 40 atom pseudocubic
BiPbZnNbO$_6$ supercell after relaxation. The $B$-site -- O bonds are
shown. The O ions are shown by (dark) red, Zn by (medium) gray,
Nb by smaller (medium) blue, Pb by gray and Bi by (smaller, light) pink
spheres. For clarity eight (2x2x2) supercells are shown. The top
panel shows a view along (001) and the bottom shows a view along (011).
Note that in addition to the cation off-centerings, there are noticeable
octahedral tilts.}
\label{struct-bipb}
\end{figure}

We started by assuming a pseudocubic structure, i.e. setting the lattice
parameters of our 40 atom supercells to be orthogonal and equal. We then
fully relaxed the internal coordinates of all atoms in the cell, with no
symmetry constraints. This was done as a function of the lattice
parameter to obtain the equilibrium cell volumes for BiPbZnNbO$_6$
and BiSrZnNbO$_6$. This was done in the LDA using the LAPW method.
The effective perovskite lattice parameters for the minimum energy
were 3.99 \AA, for BiPbZnNbO$_6$ and
3.97 \AA, for BiSrZnNbO$_6$.
As mentioned, (Bi$_{0.5}$Sr$_{0.5}$)(Zn$_{0.5}$Nb$_{0.5}$)O$_3$
was synthesized by Kosyachenko and co-workers. \cite{kosyachenko}
They reported a weakly tetragonal structure with lattice parameters
$a$=$b$=4.002, and $c$=4.015 \AA ~($c/a$=1.003) at 293 K.
These values are $\sim$ 0.9 \% larger than our calculated LDA lattice
parameter. This size of underestimation is typical of LDA errors.

\begin{figure}
\includegraphics[height=0.99\columnwidth,angle=270]{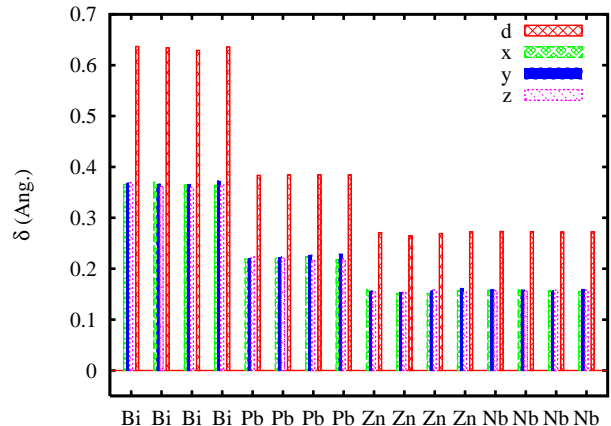}
\caption{(color online) Cation off-centerings
along the Cartesian directions with respect to their
O cages for the various sites in LDA relaxed structure of the pseudocubic
40 atom supercell of BiPbZnNbO$_6$ at a lattice parameter of 3.99 \AA.
The total displacement magnitude of each ion is also given.}
\label{fig:disp-N4C100}
\end{figure}

\begin{figure}
\includegraphics[height=\columnwidth,angle=270]{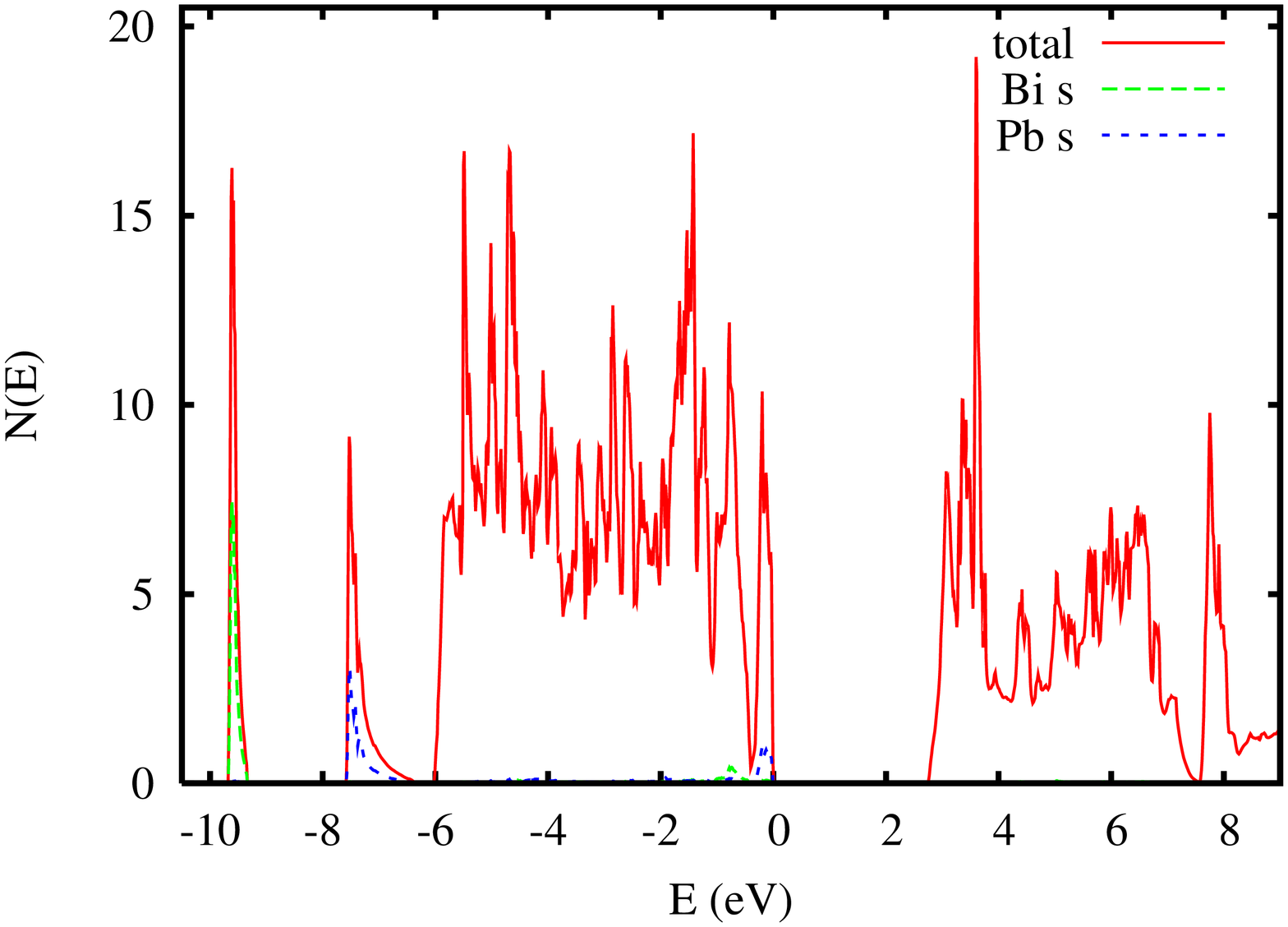}
\includegraphics[height=\columnwidth,angle=270]{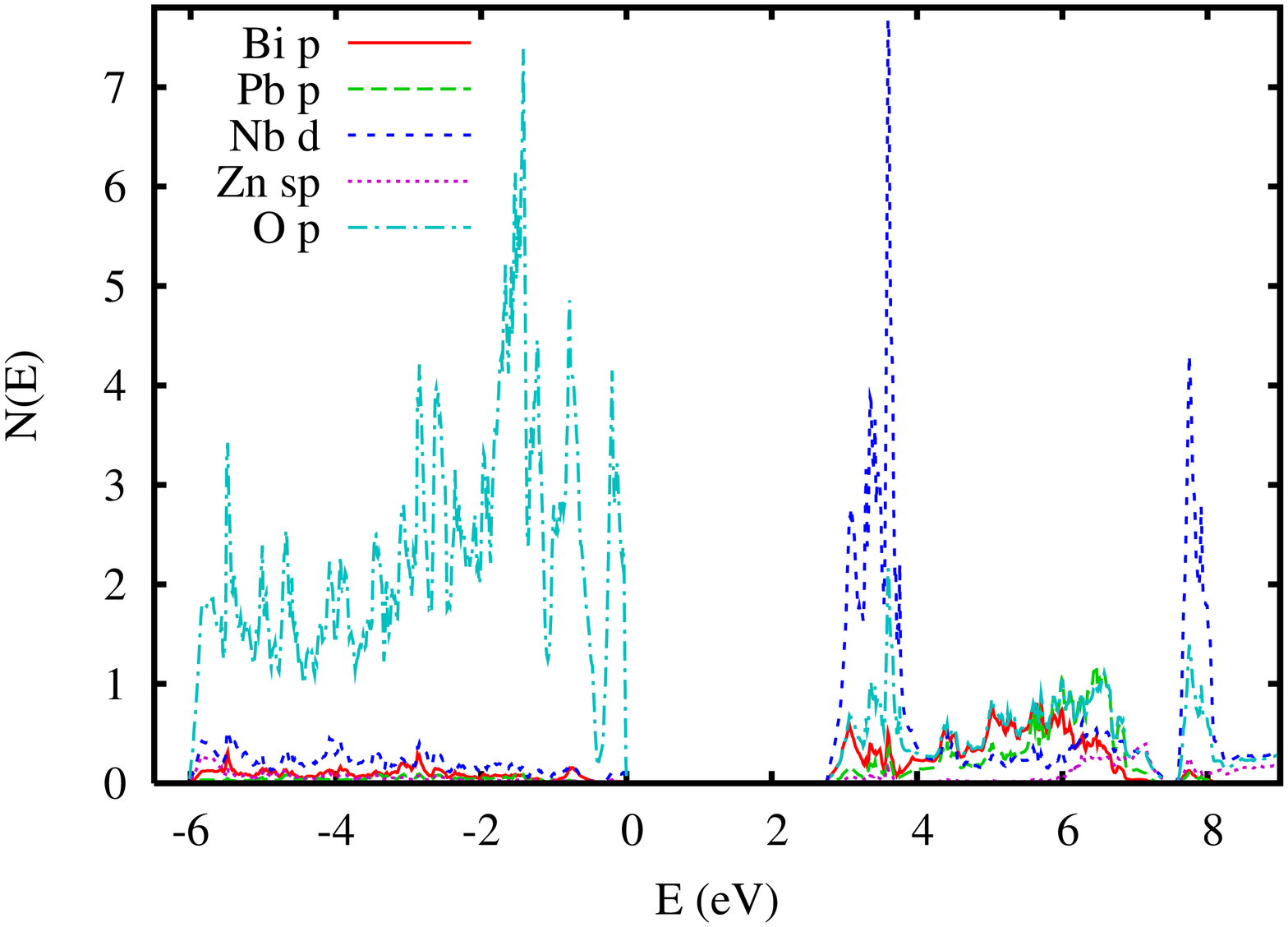}
\caption{(color online) Electronic DOS of the pseudocubic relaxed
BiPbZnNbO$_6$ supercell on a per formula unit basis. The top panel
shows the total DOS and the Bi $s$ and Pb $s$ projections. The bottom
panel shows Bi $p$, Pb $p$, Nb $d$, Zn $sp$ and O $p$ projections.
The projections are the integrals of components of the charge density
with given angular character within the LAPW spheres. Note that
for extended orbitals this gives an underestimate. The energy
zero is set at the valence band maximum.
}
\label{bipb-dos}
\end{figure}

\begin{figure}
\includegraphics[height=0.99\columnwidth,angle=270]{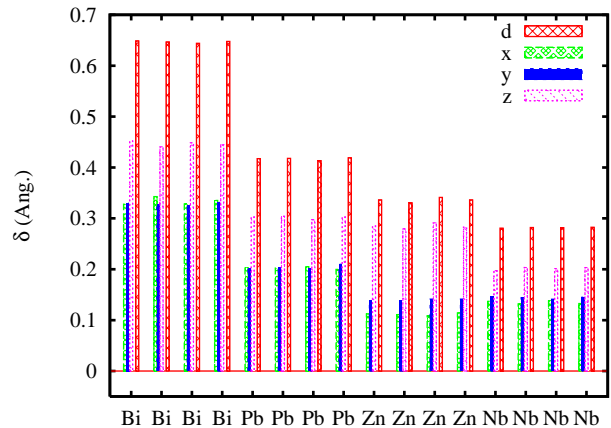}
\caption{(color online) Cation off-centerings
along the Cartesian directions with respect to their
O cages for the various sites in LDA relaxed structure of the
40 atom supercell of BiPbZnNbO$_6$ at at the same volume as in Fig. 
\ref{fig:disp-N4C100}, but with an imposed $c/a$ ratio of 1.0606.
Note that despite the large tetragonal strain the cation displacements
remain nearly collinear and
do remain reasonably close to a [111] direction.}
\label{fig:disp-N4C101}
\end{figure}

The structure of the 40 atom pseudocubic BiPbZnNbO$_6$ supercell at the LDA
equilibrium volume is depicted in Fig. \ref{struct-bipb}.
As may be seen, there are large off-centerings of the Bi ions and also
some octahedral tilting. 
Fig. \ref{fig:disp-N4C100} shows the cation off-centerings in this cell
for the different sites with respect to the centers of their O cages
(the 12 nearest O ions for the $A$-site and the 6 nearest for the $B$-site).
The largest off-centerings are of the Bi ions,
consistent with Fig. \ref{struct-bipb}, and
what might be expected based on the small size and stereochemical
activity of Bi$^{3+}$.
These displacements are large (0.64 \AA) and collinear along a [111]
direction.
Importantly, there are also large collinear displacements of the other
cations. The Pb ions displace by $\sim$ 0.39 \AA,
the Zn by $\sim$ 0.27 \AA, and
the Nb by $\sim$ 0.27 \AA, all very close to [111].
This cooperative nature of the ferroelectricity, where all cations
off-center significantly, is a characteristic of good perovskite ferroelectrics
such as KNbO$_3$, BaTiO$_3$ and PbTiO$_3$.
\cite{ghita,boyer}
These displacements
would yield a point charge polarization based on nominal charges
of 58 $\mu$C/cm$^2$ using nominal charges, with the largest
contributions coming from the Bi and Nb (note the +5 charge state of Nb).
The actual polarization is higher (see below) because the nominal charges
are enhanced due to covalency.

As is usual in perovskite ferroelectrics, we find that the ferroelectricity
is enhanced as the volume is expanded
and conversely under compression.
However, over the volume range
investigated (pseudocubic lattice parameter,
3.88 \AA$~\leq a \leq$ 4.07 \AA),
the material remains ferroelectric, with cation displacements along [111].
The average cation displacements as a function of $a$ are given in
Table \ref{tab:n4}.
The LDA typically underestimates lattice parameters,
often by $\sim$ 1--2 \%,
and as such the prediction of ferroelectricity in this material would
seem to be robust against LDA volume errors.

\begin{table}
\caption{\label{tab:n4}
Average displacements of the cations relative to their O cages
in relaxed 40 atom BiPbZnNbO$_6$ cells as a function
of pseudocubic lattice parameter, $a$.}

\begin{tabular}{|l|cccc|}
\hline
$a$(\AA)~~~~~ &
~~~$\delta_{\rm Bi}$(\AA)~~~ & ~~~$\delta_{\rm Pb}$(\AA)~~~ &
~~~$\delta_{\rm Zn}$(\AA)~~~ & ~~~$\delta_{\rm Nb}$(\AA)~~~ \\
\hline
3.88 & 0.50 & 0.28 & 0.20 & 0.22 \\
3.92 & 0.54 & 0.31 & 0.23 & 0.24 \\
3.98 & 0.62 & 0.38 & 0.26 & 0.27 \\
3.99 & 0.64 & 0.39 & 0.27 & 0.27 \\
4.03 & 0.69 & 0.44 & 0.30 & 0.30 \\
4.07 & 0.75 & 0.50 & 0.35 & 0.32 \\
\hline
\end{tabular}
\end{table}

We now discuss the electronic structure as it relates to ferroelectricity.
The electronic density of states (DOS) and projections for the relaxed
supercell are shown in Fig. \ref{bipb-dos}.
The calculated LDA band gap is 2.7 eV.
This is almost certainly an underestimate due to the LDA band gap error.
Considering the sizable gap and the fact that all the ions are in
chemically stable valence states it is likely that BiPbZnNbO$_6$ can
be made as a good insulator at ambient temperature.
This is an important consideration for a ferroelectric material to be
used in applications.
The stereochemical activity of Bi$^{3+}$ and Pb$^{2+}$ is associated
with the lone pair chemistry, which is usually discussed in terms of
a high polarizability of the electrons in the occupied $6s$ orbitals of these
ions.
As shown, the Bi $6s$ states give rise to the DOS peak centered at -9.5 eV
relative to the valence band maximum (VBM), while the Pb $6s$ states are
at -7.4 eV.
The O $2p$ bands provide the DOS from -6 eV to the VBM,
while the conduction bands are derived from unoccupied orbitals
of metal character, primarily Nb $4d$, Bi $6p$, Pb $6p$ and 
at higher energies, Zn $sp$ character.
There is a modest hybridization between the  Bi and Pb $6s$ states
and the O $2p$ states as may be seen from the Bi and Pb $s$ character
at the top of the O $2p$ bands. However, this is a mixing of
occupied states and as such does not signify chemical bonding.
Instead, as has been noted previously for ferroelectric perovskites,
\cite{cohen1992nature,zhong}
the important hybridization is between the O $2p$ states and
unoccupied metal states, which favors ferroelectricity due to chemical
bonding and leads to enhanced Born effective charges.
The energy position of the Nb $4d$ states,
which are dominant in the lower conduction bands from the
conduction band minimum (CBM) to $\sim$ 1.2 eV above it is important
because the low energy favors hybridization with O $2p$ states
and ferroelectric off-centering, as was discussed in the
case of PMN and KNbO$_3$ in relation to KTaO$_3$.
\cite{suewattana,djs-kt}
In the present case, there is a strong mixing of
O $2p$ and Nb $4d$ states.
This is evident from the Nb $d$ character in the O $2p$ bands.
The Bi $6p$ and Pb $6p$ states, which occur in the conduction
bands from the CBM to $\sim$ 4 eV above, also mix strongly with
the O $2p$ bands. There is also cross-gap hybridization evident
involving the Zn $sp$ states, although this mixing is weaker
perhaps because the Zn states occur further above the CBM.
The values of the Born effective charges are enhanced by this
cross gap hybridization. The average values for Bi, Pb, Zn and Nb
are 4.4, 3.6, 2.9 and 5.8, respectively.
The calculated polarization for the supercell is 85 $\mu$C/cm$^{2}$.
This number was obtained using a Berry's phase calculation.
The estimate of the polarization
that is obtained by multiplying the Born charges and
cation off-centerings in their O cages is  83 $\mu$C/cm$^{2}$ in close
agreement with the Berry's phase calculation.
For comparison the polarization of BiFeO$_3$ from DFT
calculations is 90 -- 100 $\mu$C/cm$^{2}$. \cite{neaton}

\begin{figure}
\includegraphics[height=0.99\columnwidth,angle=270]{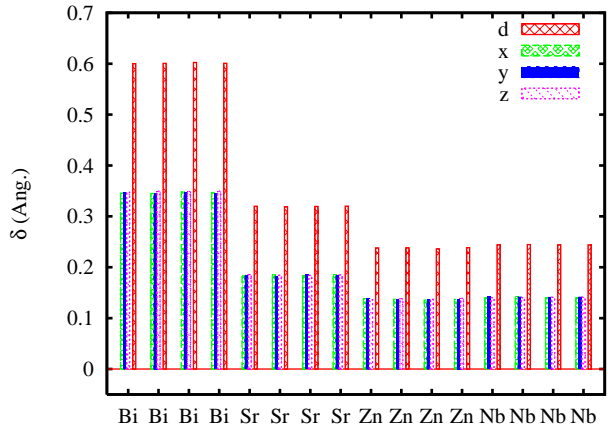}
\caption{(color online) Cation off-centerings
along the Cartesian directions with respect to their
O cages for the various sites in LDA relaxed structure of the pseudocubic
40 atom supercell of BiSrZnNbO$_6$ at a lattice parameter of 3.97 \AA.
The total displacement magnitude of each ion is also given.
Note the reduced off-centering of Sr relative to
Pb (Fig. \ref{fig:disp-N4C100}) in BiPbZnNbO$_6$, while the off-centerings
of the other cations are similar.
}
\label{fig:disp-NSVE}
\end{figure}

The above discussion in terms of cross gap hybridization due to
interaction between O $2p$ states
and extended, low lying metal orbitals making up the
conduction bands is different from a discussion in terms of a polarizable
atomic like $s$ orbital. However, it should be emphasized that atomic
character is not a uniquely defined quantity since it is basis set
dependent (note that an expansion in terms of radial functions times
spherical harmonics about a single site is a complete basis, and so
expansion about the different sites on a lattice is necessarily overcomplete).
Furthermore,
polarization of an $s$ orbital implies a mixing in of some $p$ character
from unoccupied orbitals, and in these perovskites this is $6p$ character
that arises from conduction bands.
The distinction between the two views is that in the lone pair description
the discussion is in terms of response to crystal potential as in a discussion
of crystal field, while in the cross-gap hybridization it is structure
dependent hybridization as in ligand field.
This is analogous to the origin of crystal field splittings in transition
metal oxide compounds, which arise primarily from hybridization with
O $2p$ states and not the non-spherical crystal potential around the
metal site.

As mentioned, high piezoelectric coupling in ferroelectric perovskites
is generally found near an MPB, most commonly between a pseudocubic
phase and a tetragonal phase, as in PZT.
We did calculations as a function of imposed tetragonal strain for
our supercell at a fixed volume equal to the equilibrium volume for
the pseudocubic cell ($a$=3.99 \AA).
Over the range considered (0.992 $< c/a <$ 1.076), the cation off-centerings
stayed collinear and primarily along a [111] direction, and in particular,
at least up to the maximum strain considered, did not switch to an [001]
average direction. As an
example, cation off-centerings for $c/a = 1.0606$ are shown in
Fig. \ref{fig:disp-N4C101}.
The lowest energy is at $c/a \sim 1.015$. However, near this minimum
the cation displacements remain very close to a [111] direction.
Instead it seems that the deviation of $c/a$ from unity is due to the
octahedral tilts and not strain coupling for the tetragonal ferroelectric
state. Since the tilts are expected to be coupled to the $A$-site cation
ordering, this may very well be an artifact of the specific ordering
used in the necessarily small supercell studied here
(note that the $A$-site cations would most likely be disordered
in this material).
In any case,
BiPbZnNbO$_6$ is most likely pseudocubic with a rhombohedral
ferroelectric state or weakly tetragonal.
As such it may be of interest to investigate the BiPbZnNbO$_6$ -- PbTiO$_3$
solid solution to determine whether an MPB is present.
We note that the tetragonality may be particularly sensitive to volume,
as is the case in PbTiO$_3$, meaning that the LDA volume error may be
particularly important for this, and also that errors due to
the use of a small ordered supercell may be particularly important for this.

\begin{table}
\caption{\label{tab:comp}
Comparison of properties of BiPbZnNbO$_6$
and BiSrZnNbO$_6$ cells as obtained from
our LDA calculations. $a$ denotes the pseudocubic
lattice parameter, the $z^*$ are the average Born
effective charges, $\delta$ are the average
cation off-centerings with respect to the O cages,
and $P$ is the polarization.}

\begin{tabular}{lcc}
\hline
~~~~~~~~~~~~~~~~~~ & ~BiPbZnNbO$_6$~ & ~BiSrZnNbO$_6$~ \\
\hline
$a$(LDA)                & 3.99 \AA & 3.97 \AA \\
$\delta$(Bi)            & 0.64 \AA & 0.62 \AA \\
$z^*$(Bi)               & 4.4      & 4.4 \\
$\delta$(Pb/Sr)         & 0.39 \AA & 0.33 \AA \\
$z^*$(Pb/Sr)            & 3.6      & 2.6\\
$\delta$(Zn)            & 0.27 \AA & 0.25 \AA \\
$z^*$(Zn)               & 2.9      & 2.8\\
$\delta$(Nb)            & 0.27 \AA & 0.25 \AA \\
$z^*$(Nb)               & 5.8      & 5.8      \\
$P(111)$        & 85 $\mu$C/cm$^2$ & 79 $\mu$C/cm$^2$ \\
\hline
\end{tabular}
\end{table}

BiPbZnNbO$_6$ has two $A$-site ions with stereochemical activity.
Bi shows large off-centerings while the Pb off-centerings are smaller,
and furthermore Bi carries a higher charge. This raises the question
of the role of Pb in the ferroelectricity. This question is also of
interest because Pb is considered hazardous. This fact motivates
searches for Pb-free replacements for the
commonly used Pb-based ferroelectrics such as PZT.
We performed calculations for 40 atom BiSrZnNbO$_6$ supercells, similar
to those discussed above for the Pb system.
Some properties are compared with the Pb system in Table \ref{tab:comp}.
We found a slightly smaller equilibrium volume for the Sr system, corresponding
to a pseudocubic $a$=3.97 \AA.
As shown in Fig. \ref{fig:disp-NSVE}, the material is ferroelectric, with
large collinear cation off-centerings along [111].
The calculated polarization is 79 $\mu$C/cm$^{2}$.
These off-centerings are similar to those
for the Pb system, except on the Sr site. This is also the case
for calculations at the same volume.
For a pseudocubic $a$=3.98 \AA, the calculated off-centerings
are 0.62 \AA, 0.33 \AA, 0.25 \AA, and 0.25 \AA, for Bi, Sr, Zn and Nb,
respectively,
as compared with
0.62 \AA, 0.38 \AA, 0.26 \AA, and 0.27 \AA, for Bi, Pb, Zn and Nb
in the Pb compound at the same volume.
This implies only slightly inferior ferroelectric properties for the Sr
compound as compared to the Pb compound.
The calculated average Born effective charges for BiSrZnNbO$_6$
are 4.4, 2.6, 2.8, and 5.8 for Bi, Sr, Zn, and Nb, respectively.
These are very similar to BiPbZnNbO$_6$, except that the Sr Born
charge is lower than that for Pb.
Importantly, the LDA band gap of the Sr compound, $E_g$=3.0 eV,
is slightly higher than that of the Pb compound.
This large band gap combined with the stable valences of the various
cations implies that highly insulating samples can be prepared. This
may make BiSrZnNbO$_6$ a useful alternative Pb-free ferroelectric.

\section{Summary and Discussion}

Density functional calculations for small supercells of double perovskite
BiPbZnNbO$_6$ and BiSrZnNbO$_6$ indicate that these may be useful ferroelectric
materials. They have high polarization due to the large displacement of Bi
and the displacements of Nb combined with its high charge.
They have substantial band gaps and stable chemistry consistent with
synthesis of highly insulating samples.
It will be of interest to measure properties experimentally,
and to investigate the solid solutions with PbTiO$_3$ to determine
if a morphotropic phase boundary is present.
Finally, we note that because these materials have mixtures of ions
with different charges on both the $A$- and $B$-sites, they should have
considerable chemical flexibility. While we considered BiPbZnNbO$_6$
in the double perovskite structure,
Bi$_{x}$Pb$_{1-x}$Zn$_{1/3+x/3}$Nb$_{2/3-x/3}$O$_3$ may
exist over a large range of $x$. BiPbZnNbO$_6$ is the $x$=0.5,
while $x$=0 is the known relaxor ferroelectric PZN.
To our knowledge, the $x$=1 perovskite
BiZn$_{4/3}$Nb$_{1/3}$O$_3$ has not been reported, and instead
the Bi$_2$O$_3$ -- ZnO -- Nb$_2$O$_5$ system shows Nb rich pyrochlore
phases that presumably compete with the perovskite.
\cite{levin}
As such, it may be possible to synthesize the perovskite under pressure,
similar to other low tolerance factor Bi based perovskites.
\cite{salak,belik}
In any case, the chemical flexibility implies a tunability of properties
that may be very helpful in obtaining useful ferroelectric properties
including MPBs.

\acknowledgments

We are grateful for helpful discussions with M. Fornari and M.H. Du.
This work was supported by
the Division of Materials Sciences and Engineering, Office of
Basic Energy Sciences, U.S. Department of Energy
(AS,VRC,DJS) and the Office of Naval Research (ST,DJS).

\bibliography{bipb}

\end{document}